\documentclass{PoS}
\usepackage{slashed} 
\usepackage{commath}
\usepackage{wrapfig}
\usepackage{JackCommands}

\rightline{\parbox{4cm}{\large\rm
ADP-15-41/T943\\
Edinburgh 2015/27\\
DESY 15-212\\
LTH 1065
}}

\title{Improved determination of hadron matrix elements using the variational method}

\ShortTitle{Improved Determination of Hadron Matrix Elements using the Variational Method}

\author{\speaker{J.~Dragos}$^{,a}$, R.~Horsley$^{b}$, W.~Kamleh$^{a}$, D.~B.~Leinweber$^{a}$ , Y.~Nakamura$^{c}$,
  P.~E.~L.~Rakow$^{d}$, G.~Schierholz$^{e}$, R.~D.~Young$^{a,f}$,
  J.~M.~Zanotti$^{a}$\\
        \llap{$^a$} CSSM, Department of Physics, The
        University of Adelaide, Adelaide SA 5005, Australia\\
        \llap{$^b$} School of Physics and Astronomy,
                    University of Edinburgh,
                    Edinburgh EH9 3JZ, UK \\
        \llap{$^c$} RIKEN Advanced Institute for Computational
                    Science, Kobe, Hyogo 650-0047, Japan \\
        \llap{$^d$} Theoretical Physics Division,
                    Department of Mathematical Sciences,
                    University of Liverpool,
                    Liverpool L69 3BX, UK \\
        \llap{$^e$} Deutsches Elektronen-Synchrotron DESY,
                    22603 Hamburg, Germany \\
        \llap{$^f$} CoEPP, Department of Physics, University of Adelaide,
                    Adelaide SA 5005, Australia \\
        E-mail: \email{jack.dragos@adelaide.edu.au}}

\author{CSSM/QCDSF/UKQCD Collaboration}
\abstract{The extraction of hadron form factors in lattice QCD using the standard two- and three-point correlator functions has its limitations. One of the most commonly studied sources of systematic error is excited state contamination, which occurs when correlators are contaminated with results from higher energy excitations. We apply the variational method to calculate the axial vector current \(g_{A}\) and compare the results to the more commonly used summation and two-exponential fit methods. The results demonstrate that the variational approach offers a more efficient and robust method for the determination of nucleon matrix elements.}

\FullConference{The 33rd International Symposium on Lattice Field Theory\\
		14 -18 July 2015\\
		Kobe International Conference Center, Kobe, Japan}

\begin{document}
\section{\label{Intro}Introduction}
Modern lattice QCD simulations are making significant advances towards the direct comparison with experimental results for a range of hadronic observables. Therefore the numerical simulations will require all uncertainties, both statistical and systematic, to be analysed and addressed. In this present work, we address the systematic uncertainty associated with excited-state contamination in baryon matrix elements. Excited-state contamination in baryons is a prime example where removing such effects is quite problematic, arising as a consequence of a weak signal-to-noise behaviour limiting us to the use of short source-sink time separations. In accordance with previous results we show how undertaking a variational approach can significantly reduce excited-state contamination. However, the strength of the current analysis is that we compare the variational analysis results with the previously utilised summation and two-exponential fit methods in a quantitative and systematic manner. We have considered a variety of nucleon observables, though here we focus on the nucleon axial vector current \(g_{A}\) as a test case. Looking at \(g_{A}\) we find the variational method to be superior to the summation and two-exponential fit methods. This proof of concept can and should be used in any lattice QCD calculation for hadronic observables in the future as the small cost in computational time greatly improves the robustness and precision of the result.


\section{\label{SimDet}Simulation Details}
This study is carried out on a \(32^3 \times 64\) lattice, with a pion mass of 460 MeV and lattice spacing of 0.074 fm \cite{Bietenholz:2010jr,Bietenholz:2011qq}. This ensemble corresponds to the SU(3)-symmetric point, where \(m_{u}=m_{d}=m_{s}\) with \(\kappa = 0.120900\). This simulation uses a clover action comprising of a stout smeared fermion action along with the tree-level Symanzik improved gluon action. The smeared sources undertaken in later sections are constructed using a gauge-invariant Gaussian smearing with a choice of \(\alpha=0.7\). To perform a variational method analysis, the smearing operation was repeated 32, 64 and 128 over both the creation and annihilation operators at a fixed source-sink separation of 13. Then to undertake a summation and two-exponential analysis, 32 sweeps of smearing was selected for each source-sink separation of 10, 13, 16, 19 and 22. A value of \(Z_{A} \sim 0.85\) calculated from \cite{Constantinou:2014fka} for this ensemble was used to renormalise.


\subsection{\label{CfunDef}Two-Point and Three-Point Correlation Functions}
Using two- and three-point correlators as per usual, we can construct a ratio of these correlators which has the large Euclidean time limit:
%
%
%
%
%
%
%
\be \label{eq:Rfac}
\Rfactor \xrightarrow{t\gg \tau \gg 0} \ \propto \  \FF ,
\ee
where \(\Eapp\) and \(\Ebp\) are the source and sink energies, respectively, referring to the state indices \(\alpha\) and \(\beta\), with momenta \(\pp\) and \(\p\) (\(\q \equiv \pp - \p\)). \(\Gamma\) is defined as the spin projector and \(O\) defined as the current insertion operator. We define the ``FF'' function as:
\be
\FF \equiv Tr \left\lbrace \Gamma \left(\frac{\slashed{p}' + \ma}{2\Eapp}\right) \JOq  \left(\frac{\slashed{p} + \mb}{2\Ebp}\right)  \right\rbrace ,
\ee
where \(\JOq\) is the resulting form factor combination for operator \(O\). To access the axial vector current \(g_{A}\), \(\Gamma = \frac{\gamma_{0} + I}{2}\gamma_{3}\gamma_{5}\) and \(O = \gamma_{5}\gamma_{3}\) are selected. 

\subsection{\label{CMT}Variational Method (Var)}
The variational method has shown to be quite a robust and useful tool for studying two-point correlators \cite{Engel:2010my,Edwards:2011jj,Mahbub:2010rm,Kiratidis:2015vpa,Mahbub:2013ala,Menadue:2011pd}. Recently, this approach has been extended to three-point correlators, specifically aiming to reduce the effect of excited state contamination in hadronic matrix elements \cite{Owen:2015fra,Hall:2014uca,Owen:2015gva,Owen:2012ts}.
The construction of this extension is briefly reviewed here (outline of \cite{Owen:2012ts}) resulting in improved two- and three-point correlators for a single state. The two-point correlation function for state \(\alpha\) in the variational approach is given by:

\be \label{eq:CM2pt}
\Gtwoptalpha = \sumx \Fpx Tr \left\lbrace  \Gamma \Lvac \phiaxp \phiazerobarp \Rvac \right\rbrace = \vaip \Gtwoptij \uajp .
\ee
Here the optimal interpolators \(\phiax\) \& \(\phiazerobar\) are constructed out of our basis of operators:

\be
\phiaxp = \sum_{i} \vaip \chi_{i}\lbxrb ,
\qquad
\phiazerobarp = \sum_{i} \uaip \overline{\chi}_{i}\lbzerorb ,
\ee
where \(\chi_{i} \) and \(\overline{\chi}_{i}\) are the creation and annihilation operators with basis index \(i\) and \(\vaip\) \& \(\uaip\) provide the linear combination required to isolate state \(\alpha\).
The same \(u\) \& \(\nu\) found for the two-point correlators can be used to find a three-point correlator with optimal coupling to the state:
%
%
\be 
\Gthreeptalpha = \vaipp \Gthreeptij \uajp .
\ee
The selection of two time values \(t_{0}\) and \(t_{0}+\Delta t\) provides a recurrence relationship for the two-point correlators. The left and right eigenvectors \(\vaipp\) and \(\uajp\) are obtained by solving a generalised eigenvalue problem and then used to project our matrix of two- and three-point correlators \cite{Owen:2012ts}.
%
%


\subsection{\label{SM}Summation Method}
As has been used many times in the past and in recent works \cite{Capitani:2012gj,Green:2014xba,Capitani:2015sba,Bali:2014nma}, a summation method has been suggested to reduce excited state contamination. The process is described by:

\be \label{eq:Sum}
S\left(\Gamma;0,t;0;O\right) = \sum_{\tau=\delta t}^{t-\delta t} \Rfactorzero \rightarrow c+t\left\lbrace \FFmm + \mathcal{O}\left(e^{-\Delta m t}\right)\right\rbrace ,
\ee
where \(\Delta m\) is the energy difference between the ground state mass (\(m \equiv m^{\alpha = 0}\)) and first excited state mass (\(m^{\prime} \equiv m^{\alpha = 1}\)). The slope with respect to \(t\) then isolates \(\FFmm\). We define \(\delta t \) as the number of current time values removed from the beginning and end when we sum the \(R\) function. Since the equation holds for any \(\delta t \), we can use this quantity to test the validity of the results obtained by this method.

\subsection{\label{TSF}Two-Exponential Method (2exp)}
Multi-exponential fits have also been suggested as a way of isolating excited-state contamination. While abandoned long ago in the spectroscopy community, many recent studies have attempted this in hadron matrix element calculations \cite{Capitani:2015sba,Bali:2014nma,Bhattacharya:2013ehc,Dinter:2011sg}. For comparative purposes, we also explore the use of a two-exponential fit.
The two-point and three-point correlation functions are parametrised at zero momentum by:

\be
\Gtwoztng = A_{m}e^{-m t} + A_{m^{\prime}}e^{-\left(m+\Delta m\right)t} ,
\ee
\be \label{eq:TSF}
\Gthreezt = A_{m}e^{-m t} \left\lbrace B_{0}+B_{1}\left(e^{-\Delta m \tau}+ e^{-\Delta m \left(t-\tau\right)}\right)+B_{2}e^{-\Delta m t}\right\rbrace .
\ee
The final fit parameters \( B_{0},B_{1},B_{2} \) correspond to:
\be \label{eq:TSFTerms}
B_{0} = \FFmm ,
\qquad
B_{1} = \sqrt{\frac{A_{m^{\prime}}}{A_{m}}}\FFmmp ,
\qquad
B_{2} = \frac{A_{m^{\prime}}}{A_{m}}\FFmpmp .
\ee
Note that \(B_{2}\) in Eq.~(\ref{eq:TSF}) can only be extracted if the fit has access to multiple sink times \(t\), as only varying the current time \(\tau\) cannot distinguish \(B_{0}\) from \(B_{2}\). Since we have access to multiple smearings, we can construct a combined fit over smearing-dependent \(A_{m}\) \& \(A_{m^{\prime}}\) with a common \(m\) and \(\Delta m\). The process for the two-exponential fit is to fit the two-point correlator over a sink time range in which a two-state ansatz provides a reasonable fit. Then we use these extracted parameters in the fit to the three-point correlator using a \(\tau\) range that also allows a two-state ansatz. 

\section{\label{Res} Results}

\begin{figure}[!t]
\begin{minipage}[t]{.48\textwidth}
\centering
\includegraphics[trim={4mm 5mm 3mm 3mm},clip,width=\textwidth]{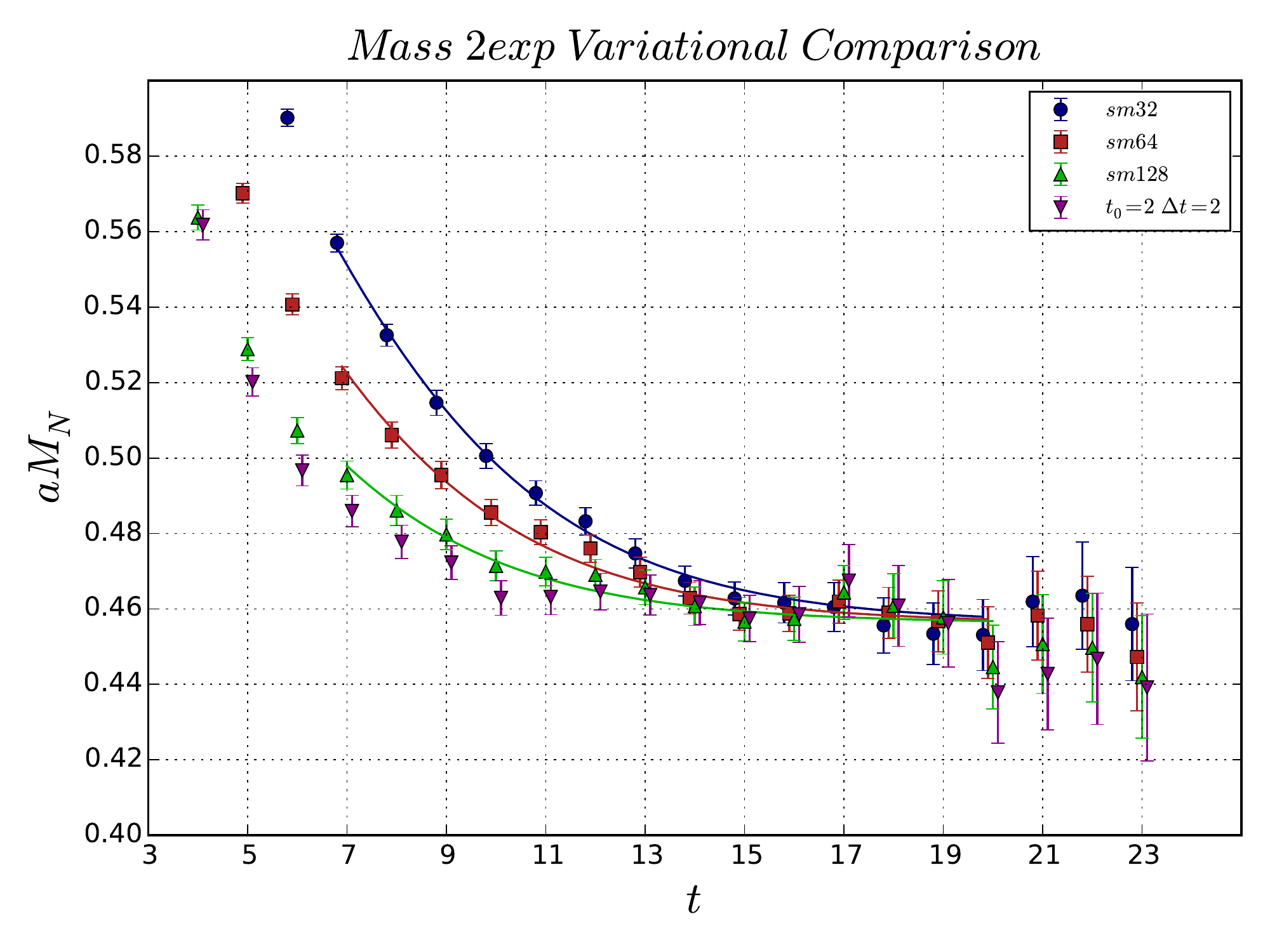}
\caption{\label{MassGraph}Effective mass plots over sink time comparing the different smearings and the variational method labelled by \(t_{0}=2\ \Delta t=2\). The lines plotted are the two-exponential fit results described in part \protect\ref{TSF}.}
\end{minipage}
\hfill
\begin{minipage}[t]{.48\textwidth}
\centering
\includegraphics[trim={4mm 5mm 3mm 3mm},clip,width=1\textwidth]{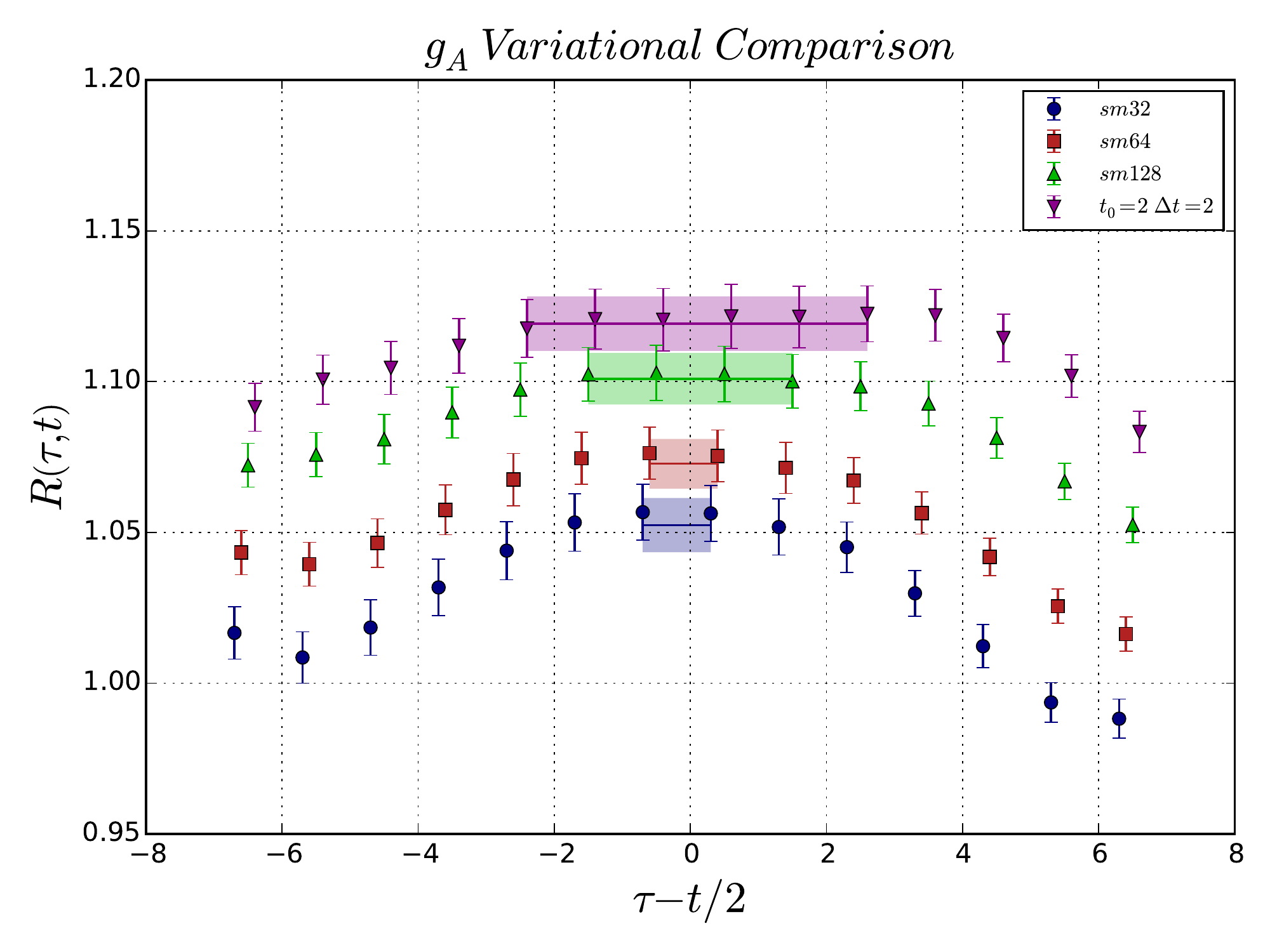}
\caption{Graph for \(g_{A}\) extracted from the R function defined in Eq.~(\protect\ref{eq:Rfac}). This plot compares individual smearing values to the variational method with a common source-sink separation of 13 lattice time slices. The lines and shaded errors correspond to the constant fit values extracted for Figure \protect\ref{gASummary}.}
\label{gAGraph}
\end{minipage}
\end{figure}

By looking at the effective mass in Figure \ref{MassGraph} we see that the variational method is producing a correlator similar to the 128 smearing-sweep result, but with reduced excited state contamination. The two-exponential fit seems to indicate that the mass plateau is slightly lower than where you might expect to get a good single state fit in the variational method, yet it is statistically consistent. This is likely an artefact of incorrectly using a single state to describe the contributions of many excited states.

In Figure \ref{gAGraph}, we see the variational method producing a superior R function compared to the individually smeared results. We also see that the variational method seems to have at least removed all the transition matrix element terms, \(B_{1}\) in Eq.~\ref{eq:TSFTerms}, as it demonstrates independence of \(\tau\) from 5 to 11. 

\begin{figure}[!t]
\begin{minipage}[t]{.48\textwidth}
\centering
\includegraphics[trim={4mm 5mm 3mm 3mm},clip,width=1\textwidth]{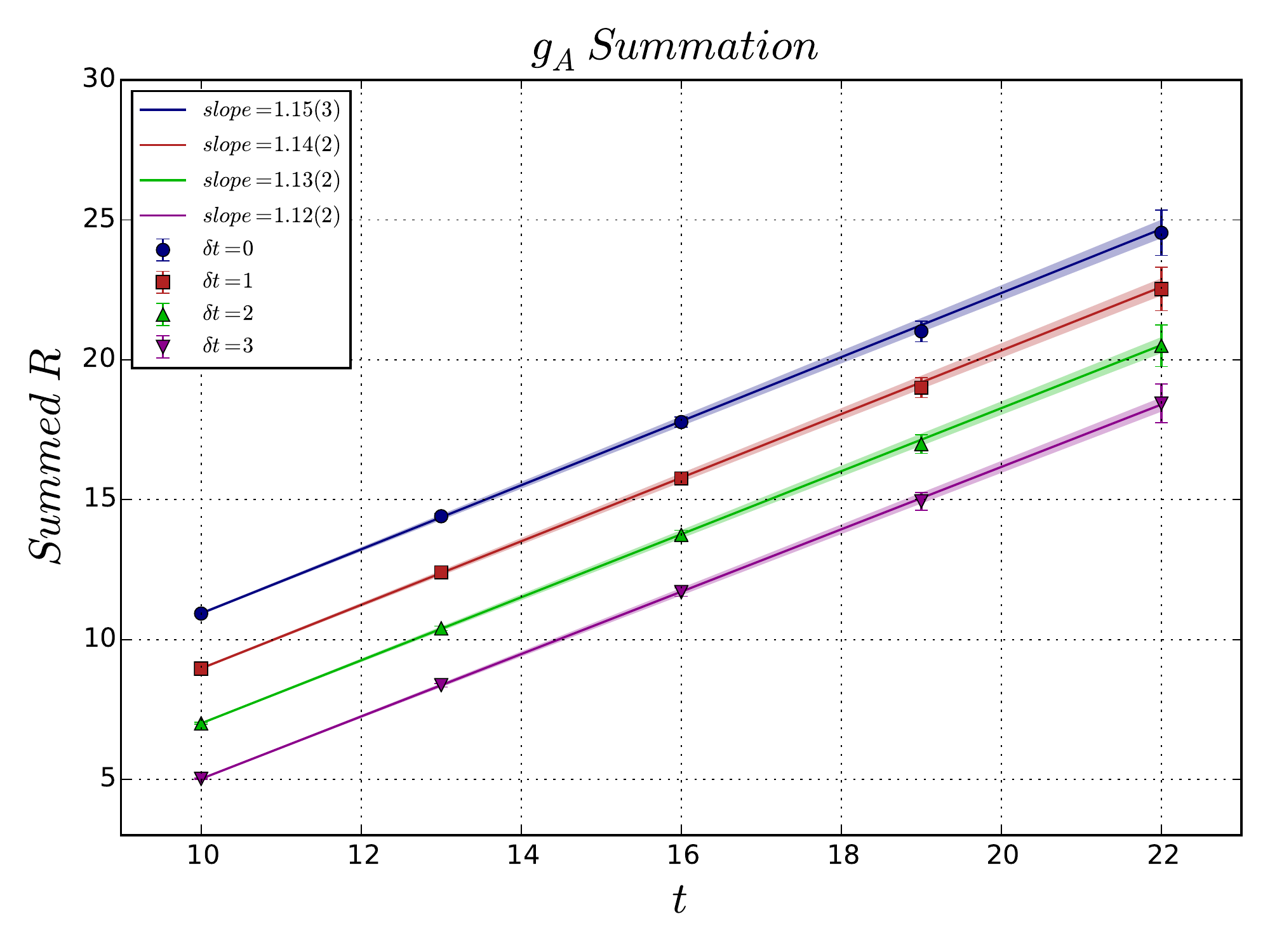}
\caption{Summed ratio factor values for multiple source/sink times (Eq.~(\protect\ref{eq:Sum})). \(\delta t\) defined in Eq.~(\protect\ref{eq:Sum}).}
\label{gASumGraph}
\end{minipage}
\hfill
\begin{minipage}[t]{.48\textwidth}
\centering
\includegraphics[trim={4mm 5mm 3mm 3mm},clip,width=1\textwidth]{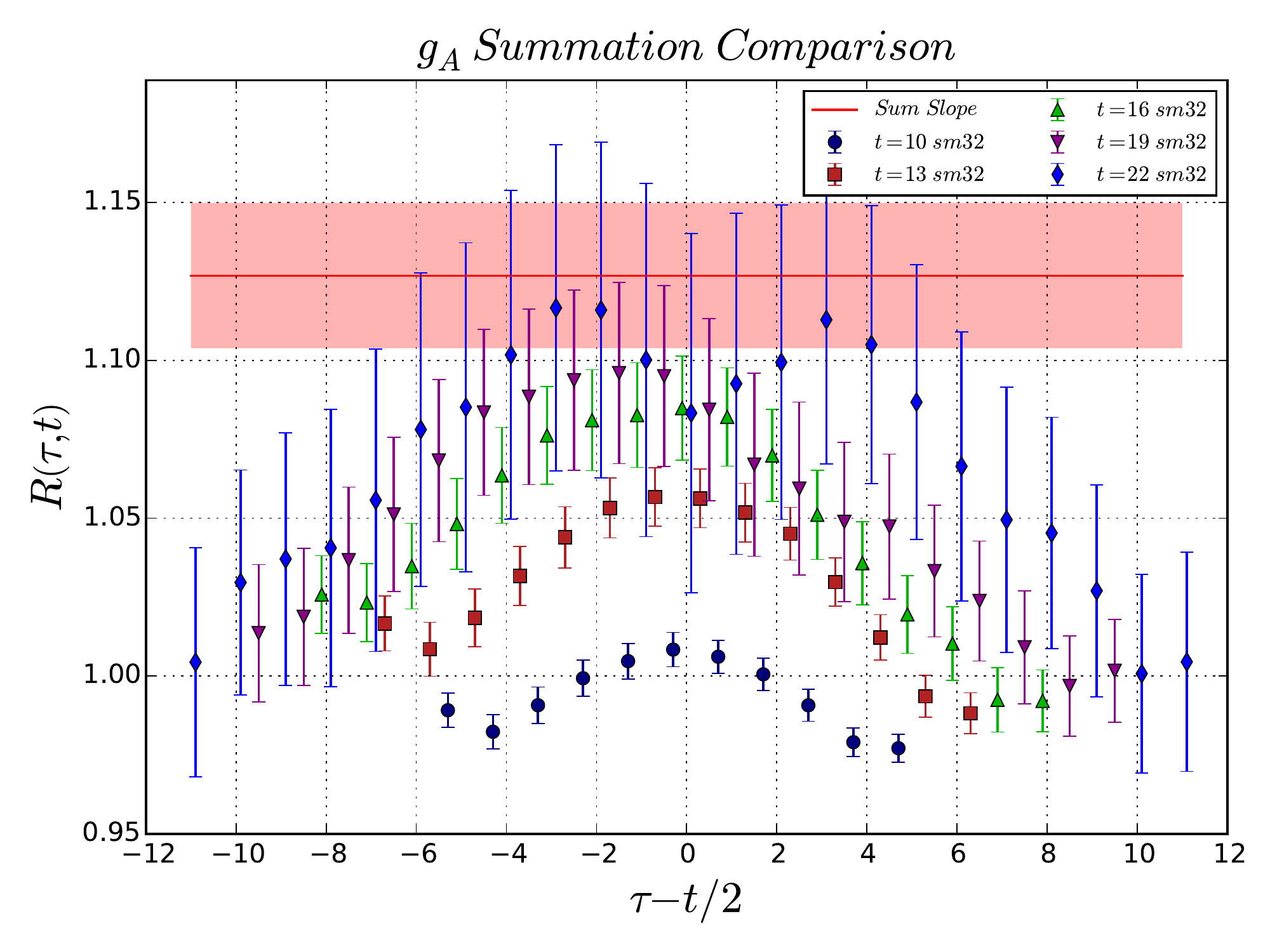}
\caption{Graph for \(g_{A}\) for multiple sink times plotted as a comparison to the summation method using \(\delta t = 3\) (purple line in Figure \protect\ref{gASumGraph}).}
\label{gASumComparisonGraph}
\end{minipage}
\end{figure}

\begin{wrapfigure}{L}{.48\textwidth}
\includegraphics[trim={4mm 5mm 3mm 3mm},clip,width=.48\textwidth]{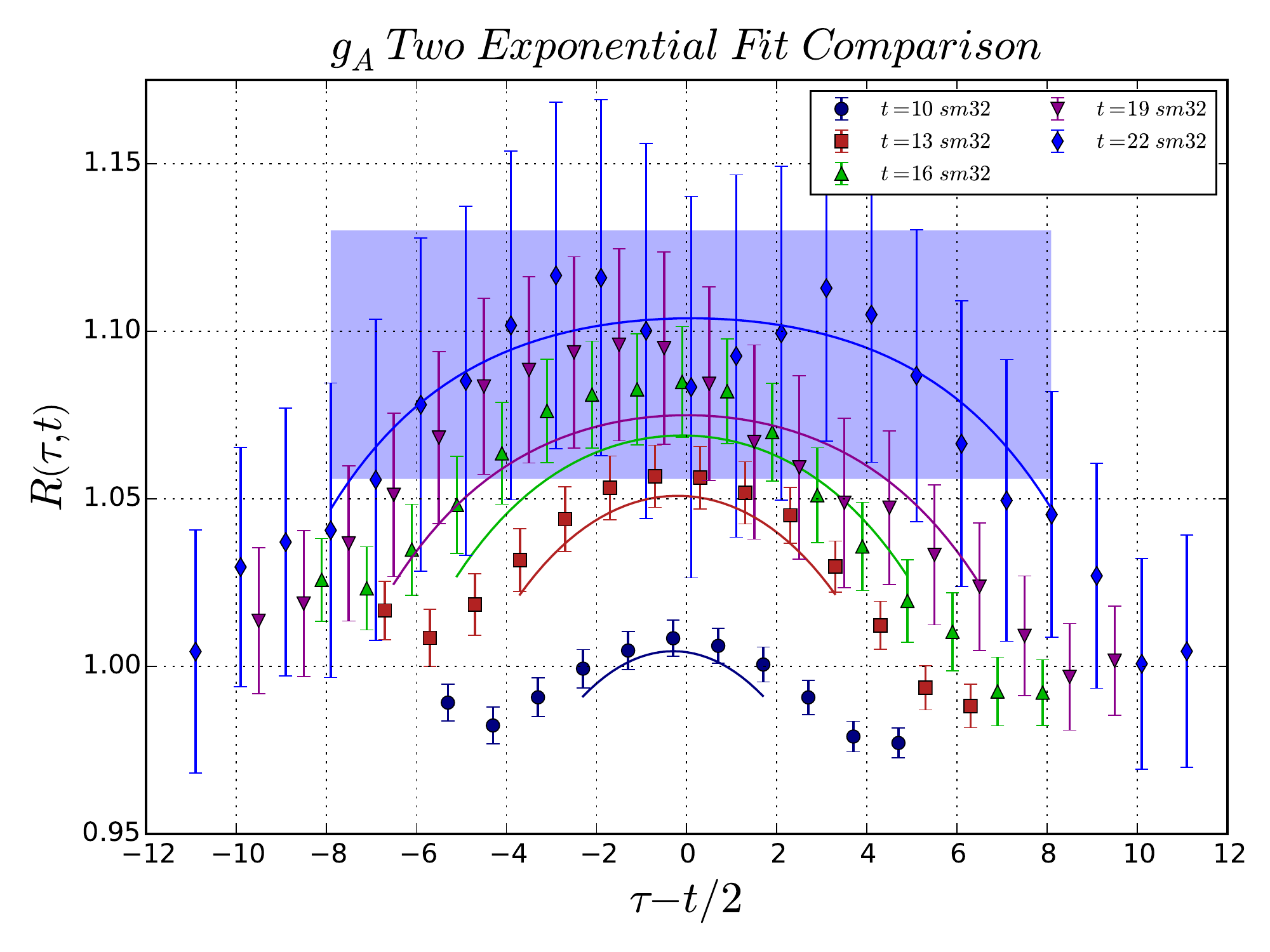}
\caption{Same as the R function values shown in Figure \protect\ref{gASumComparisonGraph}, overlaid with a two-exponential fit over both current and sink times (\(t\) \& \(\tau\)). The lines correspond to the two-exponential fit function constructed and the shaded area corresponds to the \(g_{A}\) parameter extracted from the two-exponential fit.}
\label{gA2expTsinkGraph}
\end{wrapfigure}

\begin{figure}[!t]
\centering
\includegraphics[trim={4mm 5mm 4mm 3mm},clip,width=.83\textwidth]{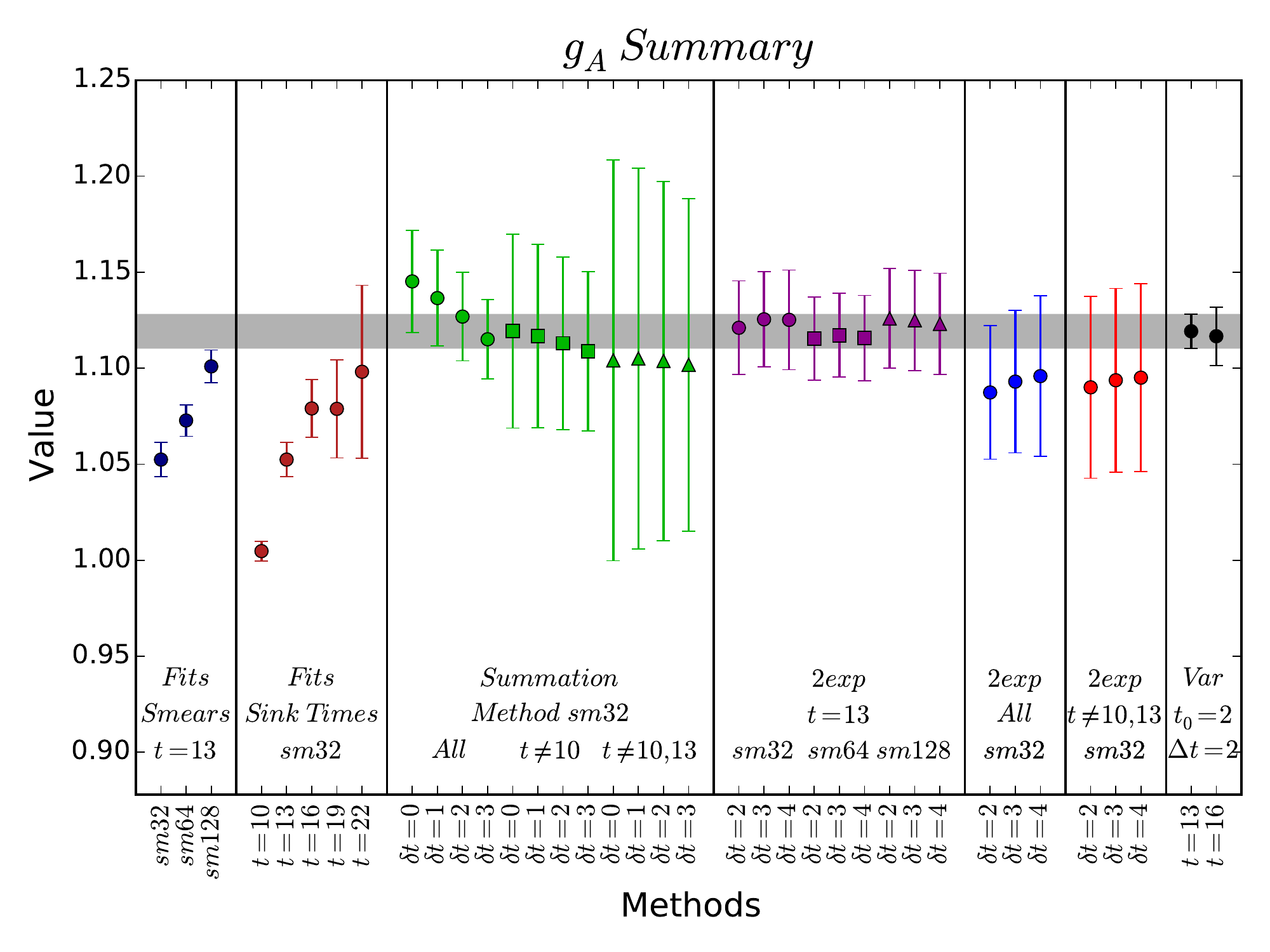}
\caption{\label{gASummary}Summary of all the extracted values for \(g_{A}\) over the different methods for source-sink separations ``\(t\)'' and smearings ``\(sm\)''. \(\delta t\) for the two-exponential method refers to the number of points excluded from the initial and final current insertion times in the analysis. See text for more details.}
\end{figure}

In Figure \ref{gASumGraph} we show the summation results for \(g_{A}\). Here we vary the total source-sink separation and compute the summed ratio defined by Eq.~(\ref{eq:Sum}). We see no statistically significant change between the 4 different calculated \(g_{A}\) values for each 4 coloured lines, which correspond to varying the \(\delta t\) value in Eq.~(\ref{eq:Sum}). We can observe that the linear fit lines are being heavily constrained by the smallest source-sink separated results due to the weighted fit, which might be problematic as these are the points that are most affected by excited state contamination. In the summary plot, Figure \ref{gASummary}, we show the results extracted when we omit the smaller source-sink separation 10 as well as omitting 10 and 13 to test how the method depends on the smaller source-sink separated result.

Turning to Figure \ref{gASumComparisonGraph}, we see that the summation result (red line in Figure \ref{gASumComparisonGraph}) produces a value which is larger than any value produced even from the largest source/sink separated points.

%

For \(g_{A}\), doing a combined two-exponential fit to all the source/sink separated data as in Figure \ref{gA2expTsinkGraph}, the result is very similar to a constant ``plateau'' fit (over a reasonable fit range) for the largest source/sink separated result. Similar to the summation method, the two-exponential method is heavily weighted by the smallest source-sink separated values which can be problematic as these values are most susceptible to excited state contamination.
%

In the final summary plot for \(g_{A}\) containing all the extracted values from all the different methods calculated (Figure \ref{gASummary}), we see a trend in the summation method to a common value when we exclude the smaller source-sink separated results (changing symbols). Applying a two-exponential fit to each of the smearings calculated at \(t=13\) (2exp / t=13 points) yields a common result that is compatible with the variational method. Excluding the smallest two source-sink separated results from the two-exponential fit (2exp \(t\ne10,13\) in Figure \ref{gA2expTsinkGraph}) we see no change to the result. 


\section{Summary}
This paper has shown how the variational approach for calculating the axial vector current has removed all statistically significant excited state contamination. This contrasts to the summation and two-exponential fit methods, where it is questionable whether the excited state contamination has been completely eliminated. The variational method undertaken in this analysis required approximately 71\% of the computing time compared to summation and two-exponential fit methods. The variational method result of \(g^{Var}_{A}=1.1203(96)\) agrees within statistical error with the Feynman-Hellmann theorem result of \(g^{FH}_{A}=1.101(24)\) \cite{Chambers:2014qaa}. Other observables not shown in this paper include the scalar charge \(g_{S}\) and the momentum fraction \(\left<x\right>\). Future work will study non-zero momentum transfers, as well as the tensor charge \(g_{T}\).


\section{Acknowledgements}
The generation of the numerical configurations was performed using the BQCD lattice QCD program, \cite{Nakamura:2010qh}, on the IBM BlueGeneQ using DIRAC 2 resources (EPCC, Edinburgh, UK), the BlueGene P and Q at NIC (J\"{u}lich, Germany) and the Cray XC30 at HLRN (The North-German Supercomputing Alliance). Some of the simulations were undertaken on the NCI National Facility in Canberra, Australia, which is supported by the Australian Commonwealth Government. We also acknowledge eResearch SA for their supercomputing support. The BlueGene codes were optimised using Bagel \cite{Boyle:2009vp}. This investigation has been supported by the Australian Research Council under grants FT120100821, FT100100005, DP150103164, DP140103067 and CE110001004.

\bibliography{skeleton}

\end{document}